\pdfoutput=1
\documentclass[iop]{emulateapj}
\slugcomment{}

\shorttitle{Unveiling molecular clouds toward bipolar HII region G8.14+0.23}
\shortauthors{L.~K. Dewangan et al.}
\begin{document}

\title{Unveiling molecular clouds toward bipolar HII region G8.14+0.23}
\author{L.~K. Dewangan\altaffilmark{1}, H. Sano\altaffilmark{2,3}, R. Enokiya\altaffilmark{3}, K. Tachihara\altaffilmark{3}, Y. Fukui\altaffilmark{2,3}, and D.~K. Ojha\altaffilmark{4}}
\email{lokeshd@prl.res.in}
\altaffiltext{1}{Physical Research Laboratory, Navrangpura, Ahmedabad - 380 009, India.}
\altaffiltext{2}{Institute for Advanced Research, Nagoya University, Furo-cho, Chikusa-ku, Nagoya 464-8601, Japan.}
\altaffiltext{3}{Department of Physics, Nagoya University, Furo-cho, Chikusa-ku, Nagoya 464-8601, Japan.}
\altaffiltext{4}{Department of Astronomy and Astrophysics, Tata Institute of Fundamental Research, Homi Bhabha Road, Mumbai 400 005, India.}
\begin{abstract}
Most recent numerical simulations suggest that bipolar H\,{\sc ii} regions, powered by O-type stars, can be formed 
at the interface of two colliding clouds. To observationally understand the birth of O-type stars, 
we present a detailed multi-wavelength analysis of an area of 1$\degr$ $\times$ 1$\degr$ hosting G8.14+0.23 H\,{\sc ii} 
region associated with an infrared bipolar nebula (BPN). Based on the radio continuum map, the H\,{\sc ii} region 
is excited by at least an O-type star, which is located toward the waist of the BPN. 
The NANTEN2 $^{13}$CO line data reveal the existence of two extended clouds at [9, 14.3] and [15.3, 23.3] km s$^{-1}$ toward the site G8.14+0.23, which are connected in the position-velocity space through a broad-bridge feature at the intermediate velocity range. 
A ``cavity/intensity-depression" feature is evident in the blueshifted cloud, 
and is spatially matched by the ``elongated redshifted cloud". 
The spatial and velocity connections of the clouds suggest their interaction in the site G8.14+0.23.
The analysis of deep near-infrared photometric data reveals the presence of clusters of infrared-excess sources, illustrating ongoing star formation activities in both the clouds. 
The O-type star is part of the embedded cluster seen in the waist of the BPN, 
which is observed toward the spatial matching zone of the cavity and the redshifted cloud. 
The observational results appear to be in reasonable agreement with 
the numerical simulations of cloud-cloud collision (CCC), suggesting that the CCC process seems to be responsible for the birth of the O-type star in G8.14+0.23. 
\end{abstract}
\section{Introduction}
\label{sec:intro}
Galactic H\,{\sc ii} regions excited by O-type stars are considered as complex physical systems, where one can expect the onset of multiple physical processes. 
Such systems are strongly affected by the intense energetic feedback of O-type stars (i.e., stellar wind, ionized emission, and radiation pressure). 
However, understanding the exact physical mechanism concerning the birth of O-type stars and young stellar clusters is extremely challenging task in the star formation research, 
and is still being debated \citep{zinnecker07,tan14,motte18}. 
The H\,{\sc ii} regions are often associated with the extended bubble/ring/bipolar infrared features \citep[e.g.,][]{churchwell06,churchwell07}. Furthermore, O-type stars may also 
trigger new groups of young stellar sources \citep[e.g.,][]{elmegreen77,bertoldi89,lefloch94,deharveng10,dewangan16a}. 

In the literature, three promising mechanisms are reported for triggering star formation in molecular clouds, which are ``globule squeezing", ``collect and collapse", and ``cloud-cloud collision (CCC)" \citep[see][for more details]{elmegreen98}.  
The first two processes are directly related to the expanding H\,{\sc ii} regions powered by massive stars, where 
new generation of star formation is triggered as the gas is swept up (and/or influenced) by the H\,{\sc ii} regions \citep[e.g.,][and references therein]{elmegreen77,whitworth94,elmegreen98,deharveng05,dale07,bisbas15,walch15,kim18,haid19}.

In the third mechanism, one expects the existence of at least two molecular cloud components in a given star-forming site, which should be connected spatial as well as in velocity space \citep[e.g.,][]{haworth15a,haworth15b,torii17,torii17x}. In the position-velocity space, the connection of two molecular clouds can be inferred through a feature at the intermediate velocity range, which is indicative of a compressed layer of gas due to the collision of the clouds \citep[e.g.,][]{torii17}. 
The basic results of the existing hydrodynamical simulations of CCC are that the CCC process has 
ability to form massive clumps and cores at the junction of molecular clouds or the shock-compressed interface layer 
\citep[e.g.,][and references therein]{habe92,anathpindika10,inoue13,takahira14,takahira18,haworth15a,haworth15b,wu15,wu17a,wu17b,torii17,balfour17,bisbas17,whitworth18}, where the initial conditions are suitable for the formation of massive stars 
and clusters of young stellar objects (YSOs). In particular, \citet{whitworth18} discussed the formation of bipolar H\,{\sc ii} regions through the CCC process. 
However, to our knowledge, observational investigations of such sites are rare in the literature.  
In this connection, the present paper is focused to observationally study a promising bipolar H\,{\sc ii} region G8.14+0.23 to explore the theoretical scenarios 
concerning the formation of bipolar H\,{\sc ii} regions hosting massive stars and embedded clusters. 

The G8.14+0.23/IRAS 17599$-$2148 is classified as a bipolar blister type H\,{\sc ii} region powered by an O6 spectral class 
source \citep[e.g.,][]{kim01,kim03}.
The G8.14+0.23 H\,{\sc ii} region is associated with a bipolar nebula (BPN) containing two mid-infrared (MIR) bubbles \citep[CN107 and CN109;][]{churchwell07}. 
Using the integrated NANTEN $^{12}$CO intensity map (beam size $\sim$2$'$.6), the site G8.14+0.23 is seen in the direction of a molecular cloud M008.2+0.2 \citep[having a radial velocity range of $\sim$12--25 km s$^{-1}$;][]{takeuchi10}. The star-forming site also hosts a Class~II 6.7 GHz methanol maser \citep[peak velocity $\sim$19.83 km s$^{-1}$;][]{szymczak12} in 
the direction of the waist of the BPN. In general, the 6.7 GHz methanol maser is considered as a good indicator of massive young stellar objects (MYSOs) 
\citep[e.g.][]{walsh98,urquhart13}. 
Using a multi-wavelength approach, \citet{dewangan12} (hereafter Paper~I) and \citet{dewangan16} (hereafter Paper~II) reported star formation activities toward the edges of the BPN as well as its waist, 
and pointed out an interaction of the H\,{\sc ii} region with its surroundings. On a large-scale study of G8.14+0.23 in Paper~II, the site has been found to be embedded 
in an elongated filamentary structure (EFS) traced in the {\it Herschel} maps. Furthermore, {\it Herschel} clumps and clusters of YSOs have also been observed toward the EFS. 
Together, the earlier studies indicate the presence of young stellar clusters and massive stars in the site G8.14+0.23. 
However, a detailed study of the distribution of molecular gas associated with the site G8.14+0.23 is yet to 
be performed. 
Figure~\ref{fg1}a shows the {\it Herschel} image at 250 $\mu$m (size $\sim$1$\degr$ $\times$ 1$\degr$) overlaid with 
the NRAO VLA Sky Survey \citep[NVSS; $\lambda$ = 21 cm; resolution = 45$\arcsec$;][]{condon98} continuum emission contours (in red). The APEX Telescope Large Area Survey of the Galaxy \citep[ATLASGAL; $\lambda$ = 870 $\mu$m; resolution = 19$''$.2; ATLASGAL;][]{schuller09} dust clumps \citep[from][]{urquhart18} 
are also marked in the {\it Herschel} 250 $\mu$m image. 
In Figure~\ref{fg1}a, sixteen clumps highlighted with square symbols (in blue) are traced in a velocity range of [15, 21] km s$^{-1}$, and at a distance of 3.0 kpc \citep[e.g.,][]{urquhart18}. 
Previously, \citet{kim01} reported a distance of 4.2 kpc to G8.14+0.23. \citet{urquhart18} made an attempt to solve the distance ambiguity of the observed ATLASGAL clumps. 
Hence, in this paper, we have considered the distance of 3.0 kpc for the site G8.14+0.23.

In order to carefully examine the large-scale molecular environment and star formation activities, we have employed new NANTEN2 $^{13}$CO (J=1$-$0) line data 
and deep UKIDSS near-infrared (NIR) photometric data. Such analysis provides a promising step for 
observationally understanding the exact birth process of massive star(s), BPN, and clusters of YSOs in the site G8.14+0.23.   

This paper is organized as follows. In Section~\ref{sec:obser}, we introduce the observational data sets adopted in this paper. 
Then, Section~\ref{sec:data} deals with the new results obtained through a multi-wavelength approach. Star formation mechanism operational in the site G8.14+0.23 is discussed in Section~\ref{sec:disc}. Finally, we conclude in Section~\ref{sec:conc}.
\section{Data sets and analysis}
\label{sec:obser}
The present paper deals with an area of $\sim$1$\degr$ $\times$ 1$\degr$ ($\sim$52.3 pc $\times$ 52.3 pc; centered at $l$ = 8$\degr$.14; $b$ = 0$\degr$.239) containing the site G8.14+0.23. 
\subsection{New Observations}
\subsubsection{$^{13}$CO (J=1$-$0) Line Data}
Observations of $^{13}$CO (J=1$-$0) emission line at 110.201353 GHz were carried out in January 2013 using 
the NANTEN2 4m millimeter/sub-millimeter radio telescope of Nagoya University installed at Pampa 
La Bola (4865m above see level), in Chile. We mapped an area of 1$\degr$ $\times$ 1$\degr$ centered at G8.14+0.23, 
using the on-the-fly mapping technique with a Nyquist sampling. 
The front-end was a 4~K cooled Nb superconductor-insulator-superconductor mixer receiver. 
The back-end was a digital Fourier-transform spectrometer with 16384 channels (1 GHz bandwidth), 
corresponding to a velocity coverage of $\sim$2700 km s$^{-1}$ and a velocity resolution of $\sim$0.17 km s$^{-1}$. 
A typical system temperature including the atmosphere is $\sim$130~K 
in the double side band. The absolute intensity was calibrated by observing Rho Oph and M17SW [($\alpha_{2000}$, $\delta_{2000}$) = (18$^{h}$ 20$^{m}$ 24.4$^{s}$, 
$-$16$\degr$ 13$'$ 17$''$.6)]. We also checked the pointing accuracy every two hours to satisfy a pointing offset within 30$''$ by observing Jupiter and IRC+10216. 
After convolution with a two dimensional Gaussian function having a full width at half maximum (FWHM) of 90$''$, we obtained the final beam size of $\sim$200$''$ (or $\sim$2.9 pc at a distance of 3.0 kpc). 
The noise fluctuation is $\sim$0.5~K at a velocity resolution of $\sim$0.5 km s$^{-1}$.
\subsection{Archival data sets}
In this paper, we also adopted multi-frequency data sets retrieved from different existing Galactic plane surveys. 
The ionized emission and the dust clumps at 870 $\mu$m are studied using the NVSS 1.4 GHz map and the ATLASGAL 870 $\mu$m data, respectively. 
The {\it Herschel} Infrared Galactic Plane Survey \citep[Hi-GAL; $\lambda$ = 70, 160, 250, 350, 500 $\mu$m; resolutions = 5$''$.8, 12$''$, 18$''$, 25$''$, 37$''$;][]{molinari10} data are utilized to study the far-infrared (FIR) and sub-millimeter (mm) emission. To explore the mid-infrared (MIR) 8.0 $\mu$m emission, the Galactic Legacy Infrared Mid-Plane Survey Extraordinaire \citep[GLIMPSE; $\lambda$ =3.6, 4.5, 5.8, 8.0 $\mu$m; resolution $\sim$2$''$;][]{benjamin03} data are used. 
To examine the sources with infrared-excess emission, the UKIRT NIR Galactic Plane Survey \citep[GPS; $\lambda$ =1.25, 1.65, 2.2 $\mu$m; resolution $\sim$0$''$.8;][]{lawrence07} and  
the Two Micron All Sky Survey \citep[2MASS; $\lambda$ =1.25, 1.65, 2.2 $\mu$m; resolution = 2$''$.5;][]{skrutskie06}) data are employed. 
More information of these selected data sets can be found in \citet{dewangan17} and \citet{dewangan17a} .
\section{Results}
\label{sec:data}
\subsection{Large-scale spatial view of G8.14+0.23}
\label{subsec:dstmas}
\begin{figure*}
\epsscale{0.78}
\plotone{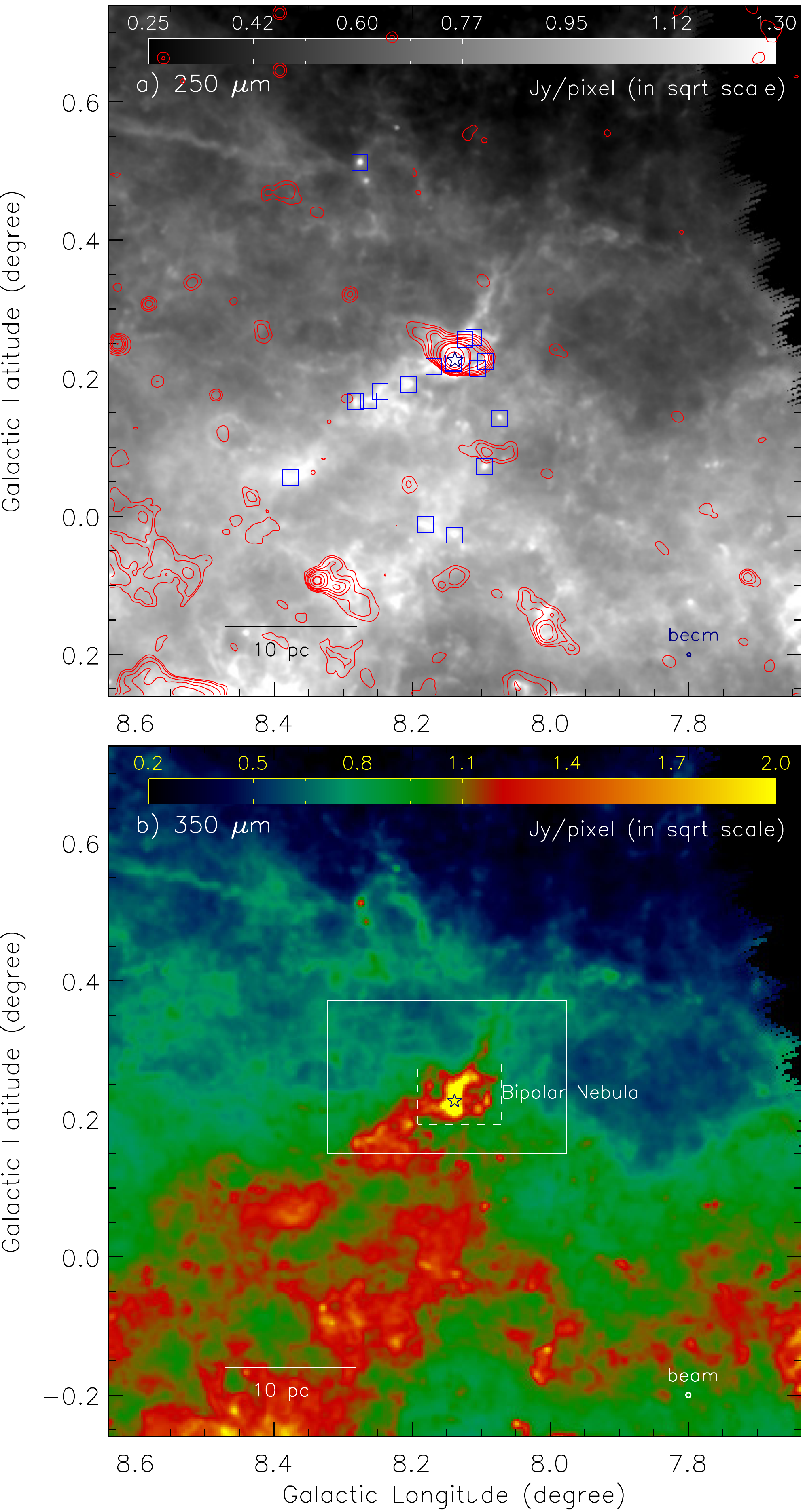}
\caption{a) Overlay of the NVSS 1.4 GHz radio continuum emission contours (in red) and sixteen ATLASGAL 870 $\mu$m dust continuum clumps \citep[blue squares; from][]{urquhart18} on the {\it Herschel} image 
at 250 $\mu$m (size $\sim$1$\degr$ $\times$ 1$\degr$; centered at $l$ = 8$\degr$.14; $b$ = 0$\degr$.24). 
The NVSS contours are shown with the levels of 0.45 mJy/beam $\times$ (8, 15, 30, 50, 80, 160, 290, 350, 900, 1800). 
All the ATLASGAL clumps are depicted in a velocity range of [15, 21] km s$^{-1}$, and are located at a distance of 3.0 kpc \citep[see][]{urquhart18}. 
b) A false color {\it Herschel} image at 350 $\mu$m. A solid box (in white) highlights the area studied in Paper~II. 
A dashed box (in white) indicates the previously known BPN (see also Figure~\ref{fg2}b). In each panel, a star symbol (in navy) indicates the position of the 6.7 GHz maser. 
The scale bar corresponding to 10 pc (at a distance of 3.0 kpc) is displayed in both the panels. In each panel, a small circle represents the beam size of the {\it Herschel} image.}
\label{fg1}
\end{figure*}

The observed sub-mm continuum and molecular line data are utilized to explore a wide-field area of the site G8.14+0.23. 
In Figure~\ref{fg1}a, the sub-mm image at 250 $\mu$m, the NVSS radio continuum emission, and the ATLASGAL dust 
continuum clumps at 870 $\mu$m are presented (see also Section~\ref{sec:intro}). 
The {\it Herschel} image at 350 $\mu$m is also presented in Figure~\ref{fg1}b. 
A qualitative information of the distribution of the cold dust emission can be gathered from the ATLASGAL and {\it Herschel} sub-mm images (see Figure~\ref{fg1} and also Section~\ref{subsec:temp} for a quantitative estimate). The previously known BPN is also seen in the {\it Herschel} images, and is highlighted by a white dashed box in Figure~\ref{fg1}b.  
In Figure~\ref{fg1}b, we have also marked an area by a solid box (in white), which was studied in Paper~II. 
The position of the 6.7 GHz methanol maser is also shown in Figures~\ref{fg1}a and~\ref{fg1}b. 
As mentioned before, the 6.7 GHz methanol maser is a reliable agent for tracing MYSOs \citep[e.g.,][]{walsh98,urquhart13}. 
The peak position of the radio continuum emission and the 6.7 GHz maser are observed toward the waist of the BPN (see also Paper~I). High resolution GMRT radio continuum maps of the G8.14+0.23 H\,{\sc ii} region were presented in Paper~II. 
Furthermore, in Paper~II, an infrared counterpart of the 6.7 GHz maser was identified with the absence of radio continuum emission, and its inner environment ($<$ 5000 AU) was also examined using the high resolution NIR images. 
Different evolutionary phases of massive stars have been reported toward the waist of the BPN (see Paper~II for more details). 

\begin{figure*}
\epsscale{0.72}
\plotone{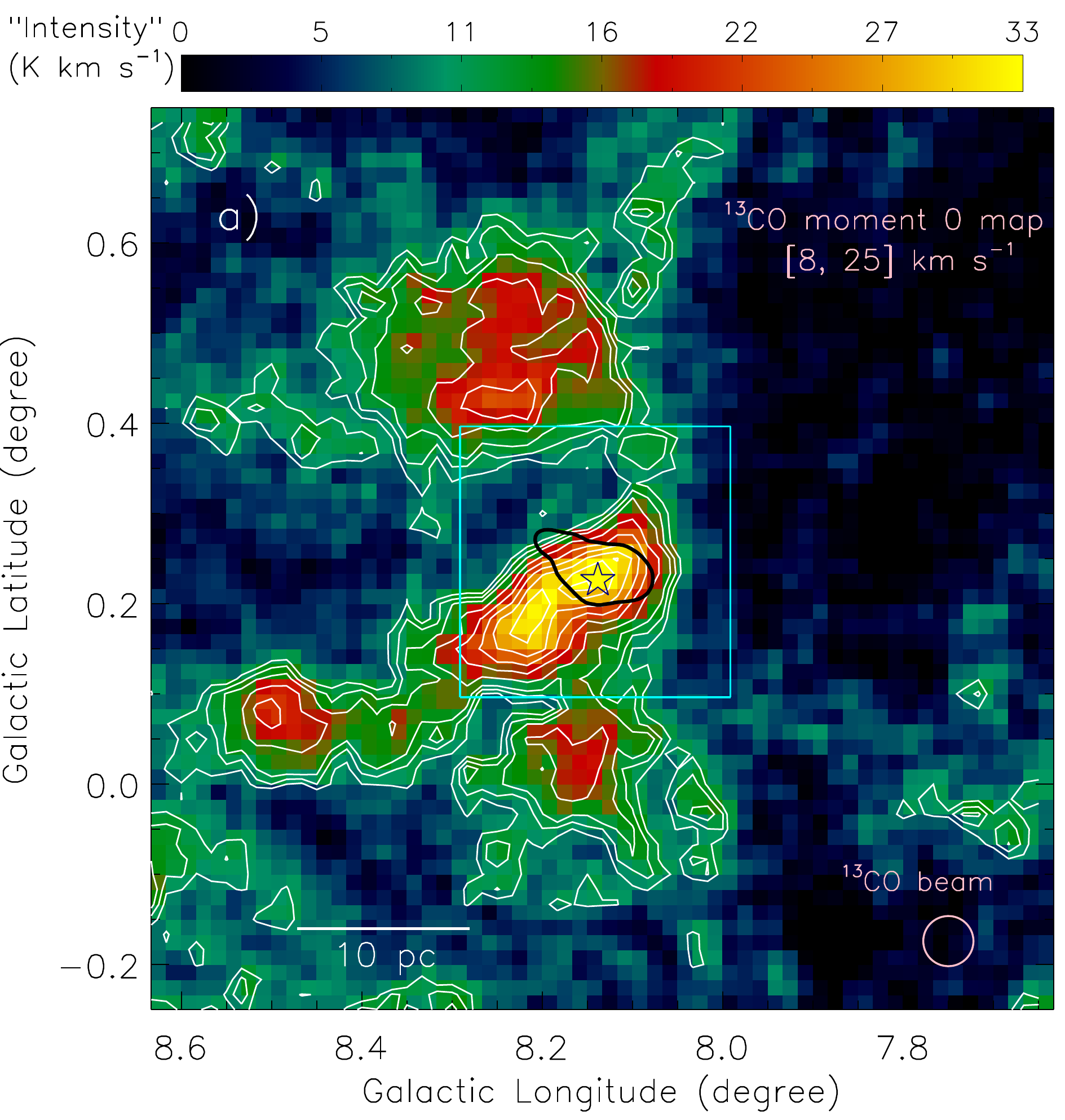}
\epsscale{0.72}
\plotone{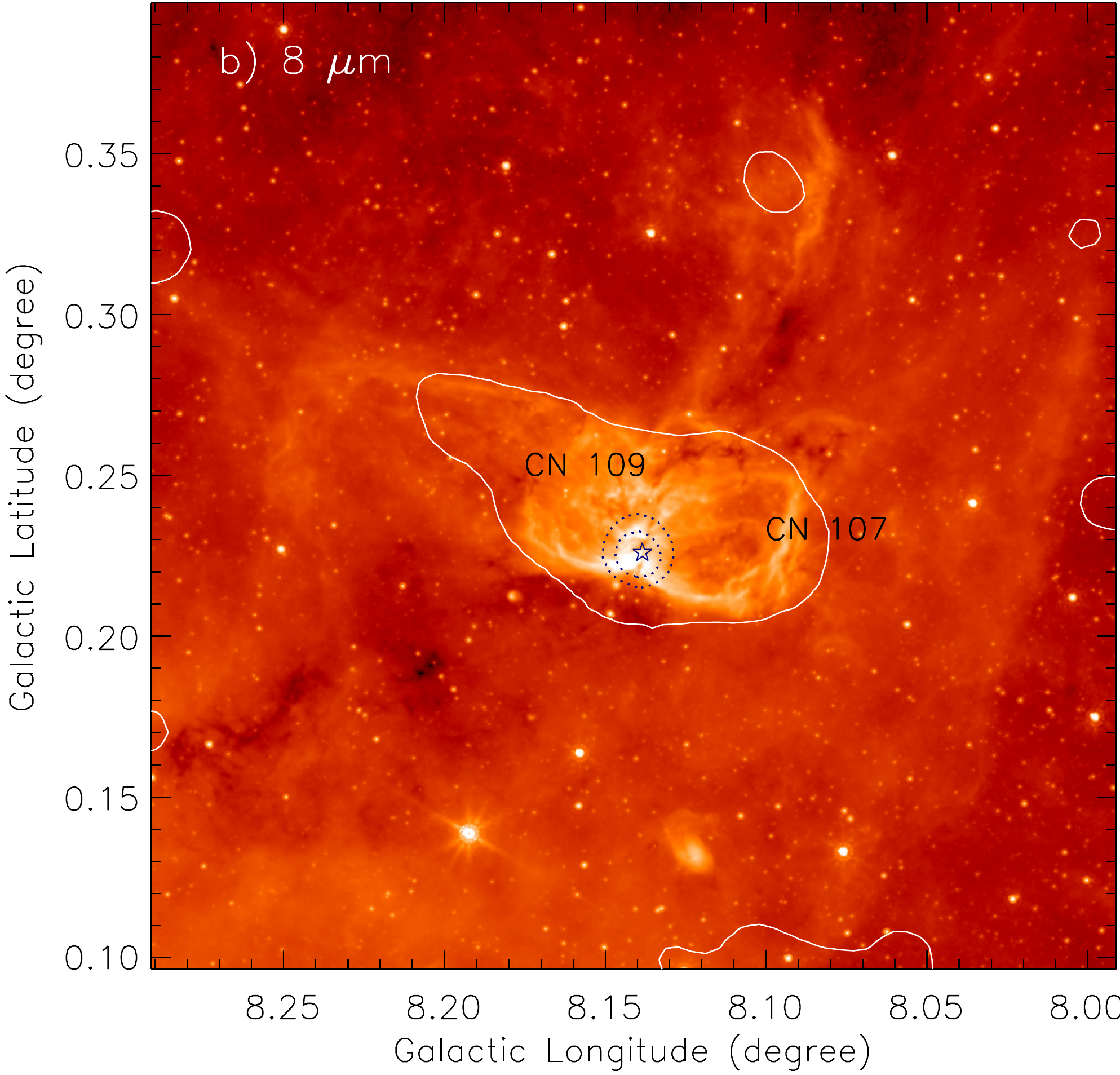}
\caption{a) The NANTEN2 $^{13}$CO(J =1$-$0) intensity map (moment-0) integrated over the velocity range of [8, 25] km s$^{-1}$. 
The map is also overlaid with the $^{13}$CO intensity contours (in white) at 34.5 K km s$^{-1}$ $\times$ (0.3, 0.35, 0.4, 0.5, 0.6, 0.7, 0.8, 0.9, and 0.98), where 1$\sigma$ $\sim$2.3 K km s$^{-1}$. The bar at the top indicates the color-coded intensity in K km s$^{-1}$. 
The NVSS 1.4 GHz radio continuum emission contour (in black) is shown with a level 
of 8$\sigma$, where 1$\sigma$ = 0.45 mJy/beam. 
A small circle (in pink) shows the beam size of the $^{13}$CO line data. b) The panel displays a false color {\it Spitzer} 8 $\mu$m image of an area highlighted by a broken box in Figure~\ref{fg2}a. 
The map is also superimposed with the NVSS 1.4 GHz radio emission contours (see white solid and black dotted contours). 
The white solid contour is the same as in Figure~\ref{fg2}a, while the black dotted contours are plotted with 
the levels of 0.81 and 1.44 Jy/beam. Two MIR bubbles CN 107 and CN 109 \citep[from][]{churchwell07} are also labeled in the image. 
In each panel, a star symbol (in navy) indicates the position of the 6.7 GHz maser.}
\label{fg2}
\end{figure*}

Using the NANTEN2 $^{13}$CO (J=1--0) line data, we have produced the $^{13}$CO intensity (moment-0) map, which is displayed in Figure~\ref{fg2}a. 
The molecular emission is integrated over a velocity range of [8, 25] km s$^{-1}$. 
The moment-0 map is also overlaid with the $^{13}$CO emission contours, revealing an extended molecular cloud in the direction of the site G8.14+0.23 (see Figure~\ref{fg2}a). Using the {\it Spitzer} 8 $\mu$m image, a zoomed-in view of the BPN is presented in Figure~\ref{fg2}b. 
The NVSS 1.4 GHz emission contour (in white) is also shown in the figure, indicating the location of 
the G8.14+0.23 H\,{\sc ii} region. 
In Paper~I, the dynamical age of the H\,{\sc ii} region powered by an O6.5-O6 star was reported to be $\sim$1.6 Myr. 
All the calculations were performed using a distance of 4.2 kpc (see Paper~I). 

Following the same methods adopted in Paper~I and considering the new distance to G8.14+0.23 (i.e., d = 3.0 kpc), we have also estimated 
the number of Lyman continuum photons (N$_{uv}$) \citep[see equation~1 in Paper~II and also][]{matsakis76} and the dynamical age of the H\,{\sc ii} region \citep[e.g.,][]{dyson80}. 
Using the NVSS 1.4 GHz data, the analysis is carried out for an electron temperature of 10$^{4}$~K and a distance of 3.0 kpc.
The analysis yields N$_{uv}$ (or logN$_{uv}$) to be $\sim$4.28 $\times$ 10$^{48}$ s$^{-1}$ (48.63) for the G8.14+0.23 H\,{\sc ii} region, 
which refers to a single ionizing star of spectral type O8-O7.5V \citep{panagia73}. 
It is indeed obvious that at least an O-type star is present toward the waist of the BPN. 

The equation of the dynamical age of the H\,{\sc ii} region \citep{dyson80} is given by\\
\begin{equation}
t_{dyn} = \left(\frac{4\,R_{s}}{7\,c_{s}}\right) \,\left[\left(\frac{R_{HII}}{R_{s}}\right)^{7/4}- 1\right] 
\end{equation}
where c$_{s}$ is the isothermal sound velocity in the ionized gas (c$_{s}$ = 11 km s$^{-1}$; \citet{bisbas09}), 
R$_{HII}$ is the radius of the H\,{\sc ii} region (R$_{HII}$ $\sim$2.3 pc), and R$_{s}$ is the radius of the Str\"{o}mgren sphere (= (3 N$_{uv}$/4$\pi n^2_{\rm{0}} \alpha_{B}$)$^{1/3}$, where 
the radiative recombination coefficient $\alpha_{B}$ =  2.6 $\times$ 10$^{-13}$ (10$^{4}$ K/T)$^{0.7}$ cm$^{3}$ s$^{-1}$ \citep{kwan97}, 
N$_{uv}$ ($\sim$4.28 $\times$ 10$^{48}$ s$^{-1}$) is defined earlier, and ``n$_{0}$'' is the initial particle number density of the ambient neutral gas).  
Adopting the values of N$_{uv}$, R$_{HII}$, and c$_{s}$, we have determined the dynamical age of the G8.14+0.23 H\,{\sc ii} region to be $\sim$0.3(1.1) Myr for 
n$_{0}$ of 10$^{3}$(10$^{4}$) cm$^{-3}$. 
\subsection{{\it Herschel} temperature and column density maps}
\label{subsec:temp}
\begin{figure*}
\epsscale{0.78}
\plotone{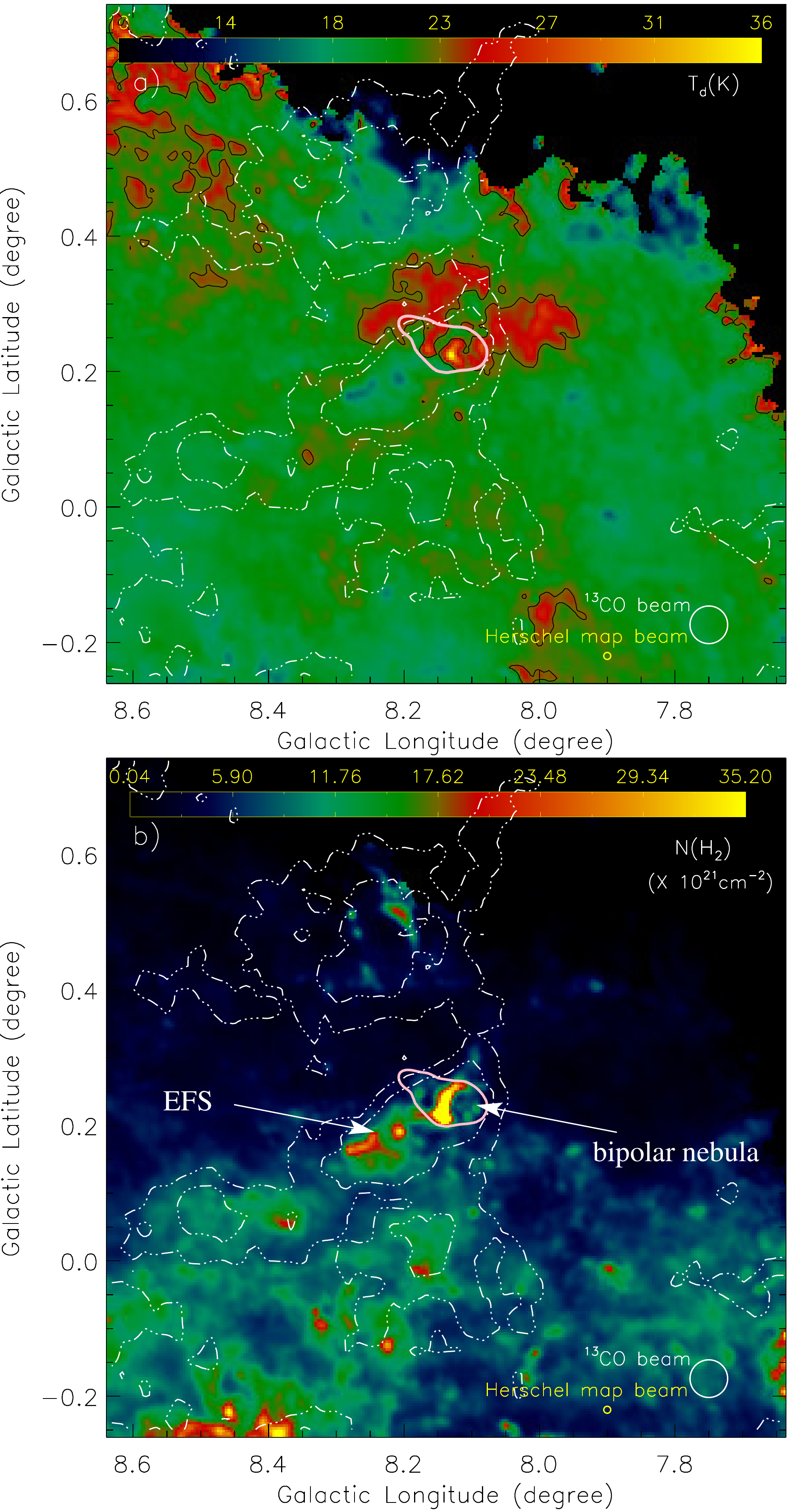}
\caption{a) The panel displays the {\it Herschel} temperature map of the G8.14+0.23 H\,{\sc ii} region.
The temperature of 23~K contour (in black) is also overlaid on the {\it Herschel} map. 
b) The panel displays the {\it Herschel} column density ($N(\mathrm H_2)$) map of the G8.14+0.23 H\,{\sc ii} region. In each panel, the integrated $^{13}$CO emission contours at [8, 25] km s$^{-1}$ (see white dotted-dashed contours) are overplotted with the levels of 10.35 (4.5$\sigma$) and 16.66 (7.2$\sigma$) K km s$^{-1}$ (see also Figure~\ref{fg2}a). 
The NVSS 1.4 GHz radio continuum emission contour (in pink) is shown with a level of 0.45 mJy/beam $\times$ 8 in both the panels. In each panel, a big circle (in white) and a small circle (in yellow) show the beam sizes of the $^{13}$CO line data and the {\it Herschel} column density map, respectively.}
\label{fg3}
\end{figure*}

Figures~\ref{fg3}a and~\ref{fg3}b present the {\it Herschel} temperature and column density maps of 
a large-scale area containing the G8.14+0.23 H\,{\sc ii} region. 
In Paper~II, we described the procedures for producing these maps \citep[see also][for more details]{mallick15}. 
The contours of the integrated intensity map of $^{13}$CO (J=1$-$0) at [8, 25] km s$^{-1}$ are also overlaid on both the 
{\it Herschel} maps (see also Figure~\ref{fg2}a). The EFS and the BPN are seen within the $^{13}$CO emission contour with a level of 16.66 K km s$^{-1}$ (see Figure~\ref{fg3}b).  
To highlight the location of the G8.14+0.23 H\,{\sc ii} region, the NVSS 1.4 GHz continuum 
emission contour is marked in the maps. 
In Figure~\ref{fg3}a, the temperature contour at 23~K is also superimposed on the {\it Herschel} temperature map, 
tracing an extended temperature feature. The observed temperature feature seems to be more extended compared to 
the radio continuum emission traced in the NVSS map, depicting the signatures of the feedback from the clusters of YSOs as well as the impact of the ionizing photons of the O-type star (see Paper~II and also Section~\ref{subsec:phot1} in this paper). In Figure~\ref{fg3}b, a comparison of the distribution of molecular 
gas against the dust column density distribution can be examined, revealing locations of high dust column density ($N(\mathrm H_2)$ $\sim$ 10--35 $\times$ 10$^{21}$ cm$^{-2}$ or $A_V$ $\sim$10.5--37.5 mag) within the cloud. 
Using the value of $N(\mathrm H_2)$, a conversion expression \citep[i.e., $A_V=1.07 \times 10^{-21}~N(\mathrm H_2)$;][]{bohlin78} is utilized to determine the extinction information along the line of sight.
\subsection{Molecular cloud components}
\label{sec:coem} 
\begin{figure*}
\epsscale{0.72}
\plotone{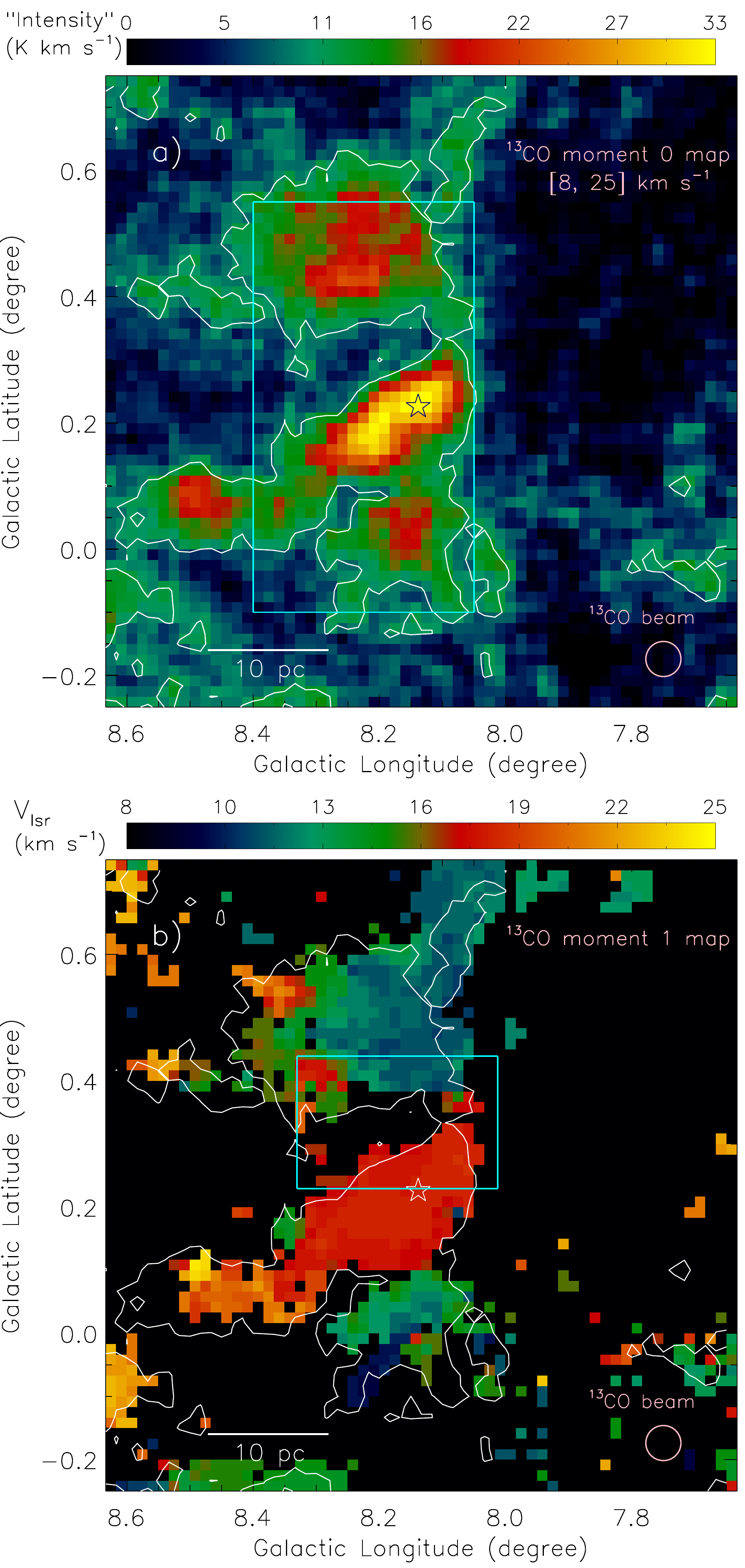}
\caption{a) $^{13}$CO intensity map (moment-0) integrated over the velocity range of [8, 25] km s$^{-1}$ (see also Figure~\ref{fg2}a). The area highlighted by a solid box (in cyan) is used for extracting the position-velocity maps (see Figures~\ref{fg5}a and~\ref{fg5}b). b) Intensity-weighted mean velocity (moment-1) map of $^{13}$CO. 
The $^{13}$CO emission integrated over the velocity range of [8, 25] km s$^{-1}$ and clipped at the 4.8$\sigma$ rms level of 
$\sim$0.44 K per channel. The area highlighted by a solid box (in cyan) is used for extracting the $^{13}$CO profile (see Figure~\ref{fg5}d). In both the panels, the $^{13}$CO emission contour at [8, 25] km s$^{-1}$ is also overlaid with a level of 10.35 K km s$^{-1}$ (4.5$\sigma$; see also Figure~\ref{fg2}a). 
In each panel, a star symbol indicates the position of the 6.7 GHz maser, and a big circle (in pink) shows the beam size of the $^{13}$CO line data.}
\label{ufg5x}
\end{figure*}

As mentioned earlier, a detailed study of molecular gas in a wide-scale environment of G8.14+0.23 is 
missing in the literature. To examine the distribution of molecular gas in our selected target field, 
Figures~\ref{ufg5x}a and~\ref{ufg5x}b display the $^{13}$CO moment-0 and moment-1 maps overlaid with the contour of the integrated intensity map of $^{13}$CO at [8, 25] km s$^{-1}$, respectively. 
In Figure~\ref{ufg5x}, one can compare the moment-0 map against the moment-1 map. 
The $^{13}$CO moment-0 map is the same as shown in Figure~\ref{fg2}a. In order to produce the moment-1 map, the $^{13}$CO emission is integrated over the velocity range of [8, 25] km s$^{-1}$, and is clipped at the 4.8$\sigma$ rms level of $\sim$0.44 K per channel. 
Figure~\ref{ufg5x}b enables us to examine the intensity-weighted mean velocity of the emitting gas, indicating the presence of 
two molecular cloud components in the direction of the site G8.14+0.23.

\begin{figure*}
\epsscale{1.2}
\plotone{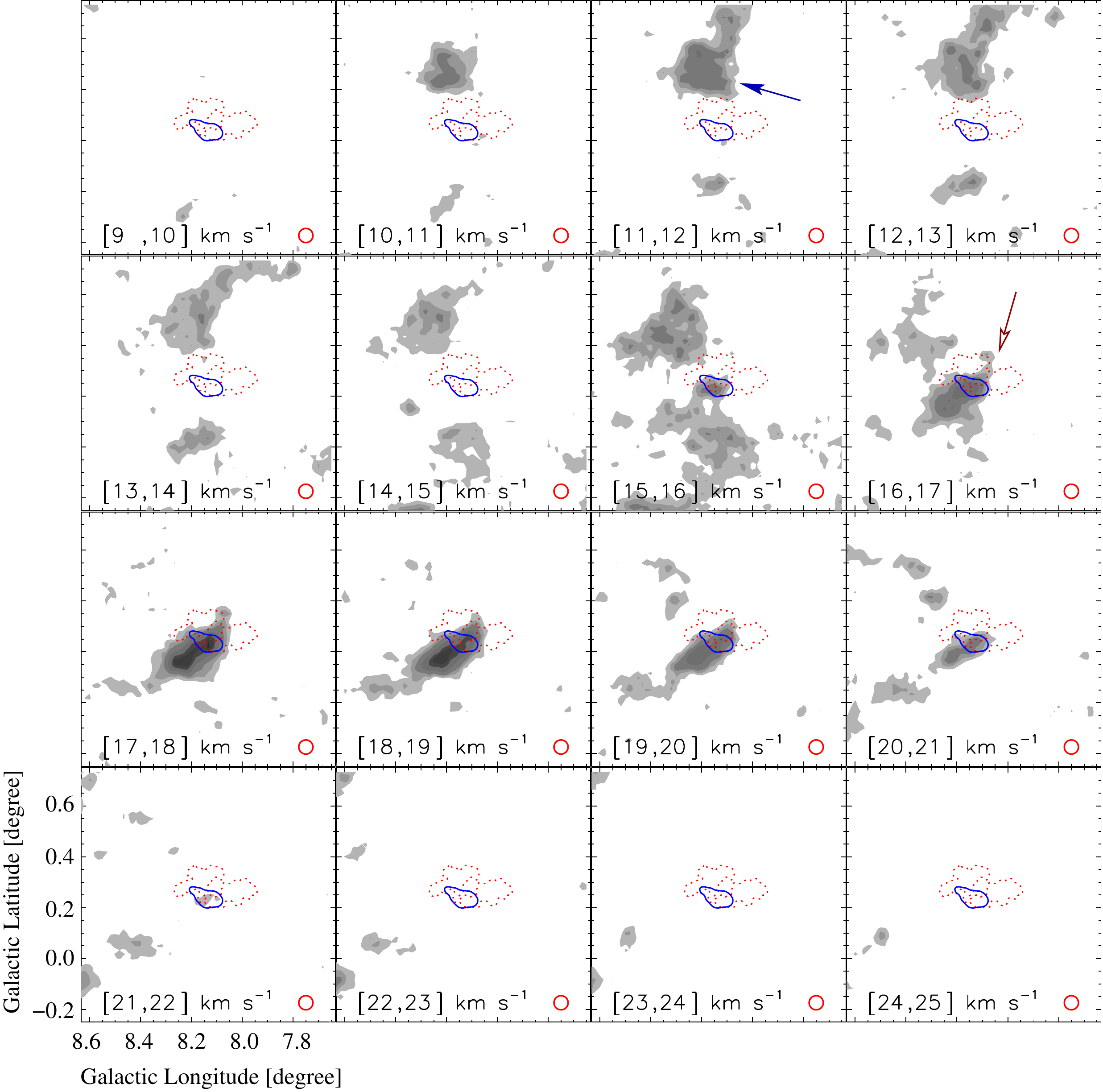}
\caption{$^{13}$CO(J =1$-$0) velocity channel contour maps.
The molecular emission is integrated over a velocity interval, which is labeled in each panel (in km s$^{-1}$). In all the panels, the contours are shown with the levels of 2.0, 3.0, 4.0, 6.0, 8.0, 9.0, and 10 K km s$^{-1}$. In all the panels, the NVSS 1.4 GHz radio continuum emission contour (in blue) toward the G8.14+0.23 H\,{\sc ii} region 
is shown with a level of 0.45 mJy/beam $\times$ 8. In each panel, the {\it Herschel} temperature contour at 23~K is also shown (see dotted red contour and also Figure~\ref{fg3}a). Arrows highlight molecular gas in two cloud components. In each panel, a thick circle (in red) shows the beam size of the $^{13}$CO line data.}
\label{fg4}
\end{figure*}

In Figure~\ref{fg4}, we display the integrated $^{13}$CO velocity channel maps from 9 to 25 km s$^{-1}$. In Figure~\ref{fg4}, each panel is 
generated by integrating the emission over 1 km s$^{-1}$ velocity intervals, and the observed temperature feature (see dotted red curve) and the location of the G8.14+0.23 H\,{\sc ii} region (see solid blue contour) are also marked. 
We find the existence of two molecular cloud components (around 11.5 and 18.5 km s$^{-1}$) 
in the direction of the site G8.14+0.23. 

\begin{figure*}
\epsscale{1.14}
\plotone{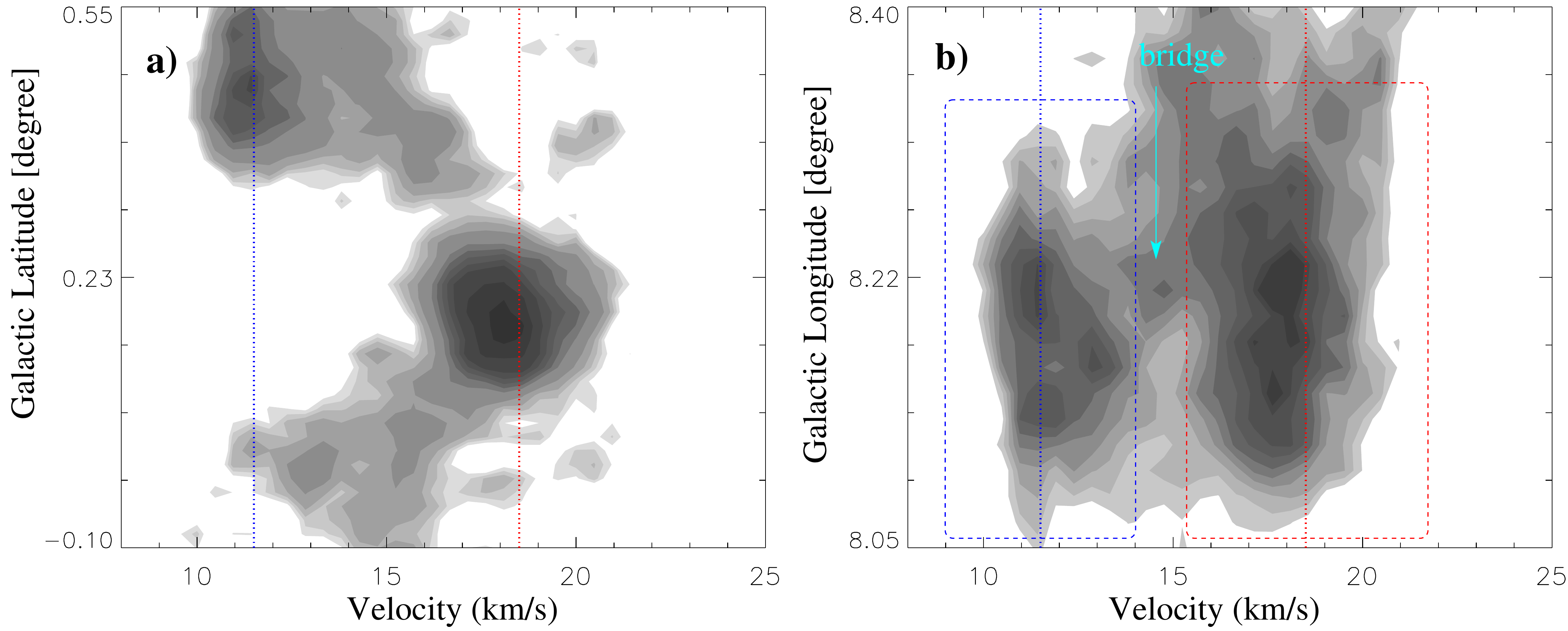}
\epsscale{0.58}
\plotone{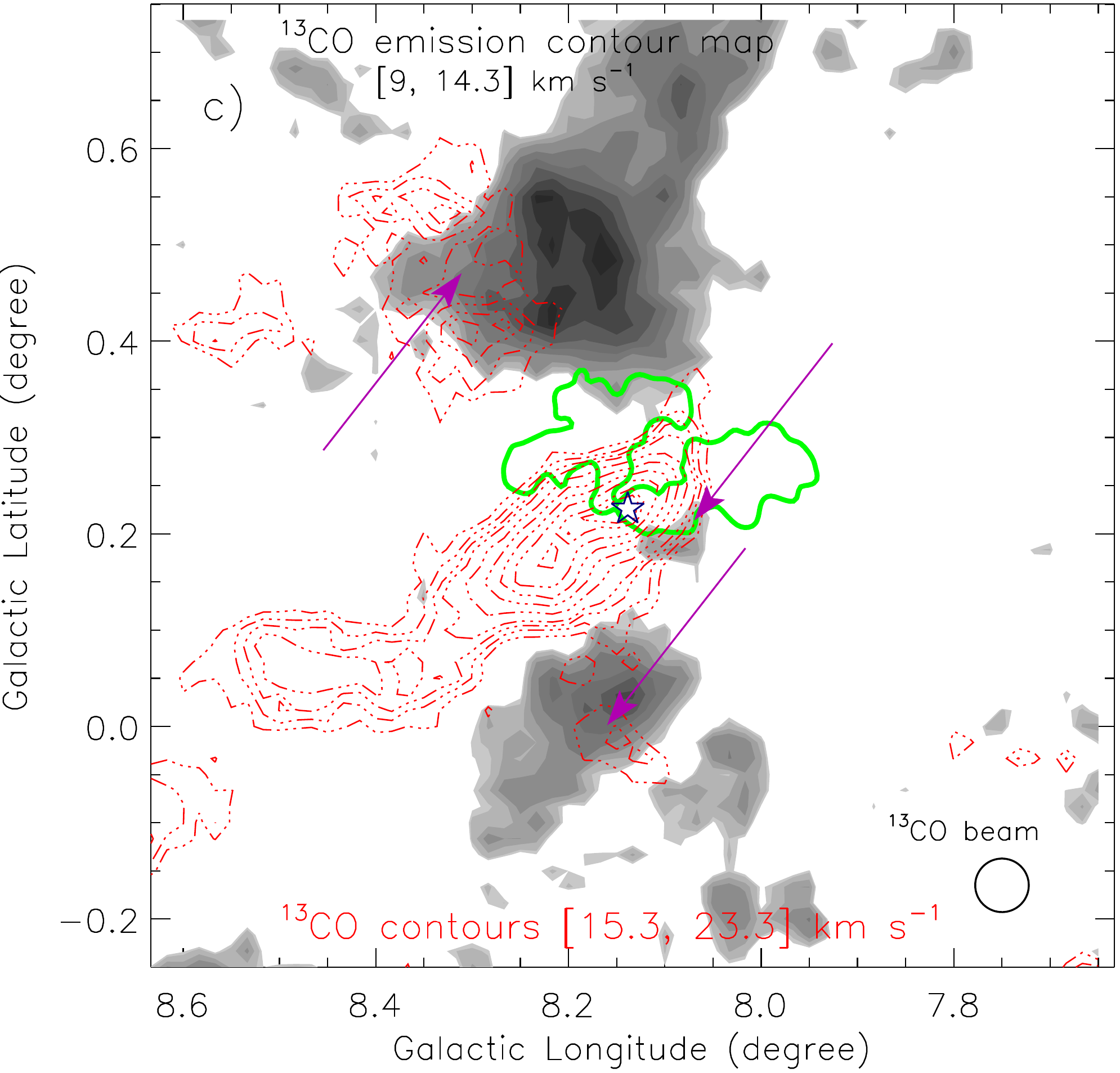}
\epsscale{0.56}
\plotone{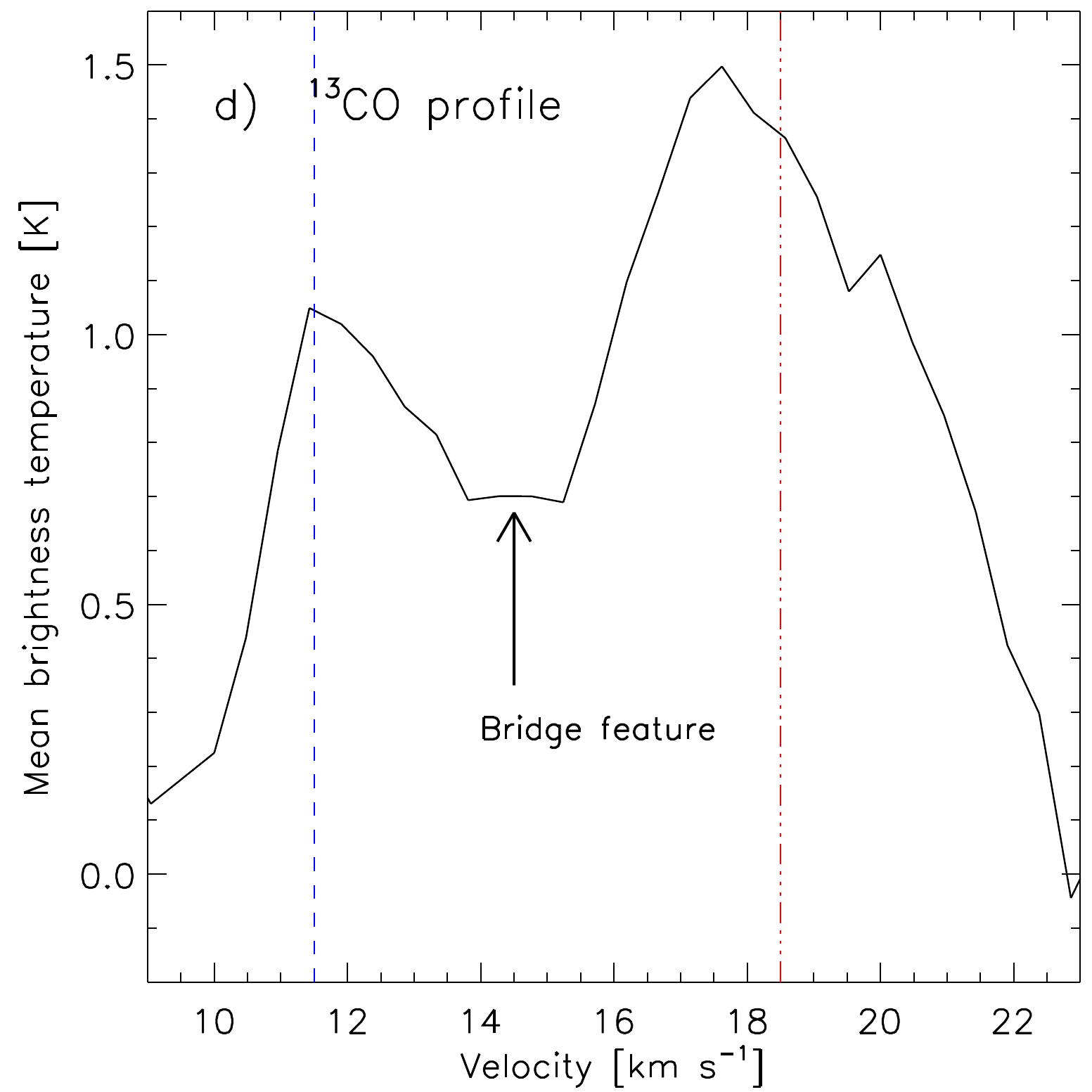}
\caption{a) Latitude-velocity map of $^{13}$CO (see a solid box in Figure~\ref{ufg5x}a). 
The $^{13}$CO emission is integrated over the longitude range from 8$\degr$.05 to 8$\degr$.40. 
The contour levels are 18, 20, 22, 25, 30, 40, 45, 50, 55, 60, 70, 80, 90, and 100 deg K, 
where 1$\sigma$ $\sim$2 deg K. 
b) Longitude-velocity map of $^{13}$CO (see a solid box in Figure~\ref{ufg5x}a). 
The $^{13}$CO emission is integrated over the latitude range from $-$0$\degr$.10 to 0$\degr$.55.
The contour levels are 30, 35, 40, 45, 50, 55, 60, 65, 70, and 72 deg K, 
where 1$\sigma$ $\sim$2.7 deg K. Two broken boxes indicate two cloud components. 
c) The $^{13}$CO emissions integrated over two different velocity ranges at [9, 14.3] 
and [15.3, 23.3] km s$^{-1}$ are presented, and 
the velocity ranges are also given in the figure. 
The contour levels of the background $^{13}$CO emission map at [9, 14.3] km s$^{-1}$ are 
3.87, 4.15, 4.84, 5.53, 6.91, 8.30, 9.68, 11.06, 12.44, and 13.14 K km s$^{-1}$, 
where 1$\sigma$ $\sim$1.2 K km s$^{-1}$. 
The broken contours tracing the cloud at [15.3, 23.3] km s$^{-1}$ (in red) are shown with the levels of 
7.87, 9.45, 11.02, 12.60, 15.75, 18.89, 22.04, 25.19, 28.34, and 29.92 K km s$^{-1}$, 
where 1$\sigma$ $\sim$1.3 K km s$^{-1}$. 
The temperature of 23~K contour (in green) is also shown in the figure. 
A star symbol indicates the position of the 6.7 GHz maser, and a big circle (in black) 
shows the beam size of the $^{13}$CO line data. 
d) The $^{13}$CO profile in the direction of an area highlighted by 
a solid box in Figure~\ref{ufg5x}b. The profile is produced by 
averaging the area shown in Figure~\ref{ufg5x}b. An arrow indicates an almost flattened spectrum between two velocity peaks. 
In panels ``a" ,``b", and ``d", two velocities (i.e., 11.5 and 18.5 km s$^{-1}$) are highlighted by two dotted lines.}
\label{fg5}
\end{figure*}

To further explore the different molecular components, we display the latitude-velocity and longitude-velocity maps of $^{13}$CO in Figures~\ref{fg5}a and~\ref{fg5}b, respectively. Note that the position-velocity maps of $^{13}$CO are extracted for an area, 
which is marked in Figure~\ref{ufg5x}a (see a solid box in Figure~\ref{ufg5x}a). In the direction of this selected area, the majority of the molecular emission is distributed. 
Both these position-velocity maps show two velocity components around 11.5 and 18.5 km s$^{-1}$ (see dotted lines in Figures~\ref{fg5}a and~\ref{fg5}b). The velocity gradients are evident in Figure~\ref{fg5}a. 
In the position-velocity space, we find a broad-bridge feature at the intermediate velocity range between two velocity components 
around 11.5 and 18.5 km s$^{-1}$ (see an arrow in Figure~\ref{fg5}b). 
In Figure~\ref{fg5}c, we have presented the spatial distribution 
of molecular gas in two clouds at [9, 14.3] and [15.3, 23.3] km s$^{-1}$. 
Very weak $^{13}$CO emission is observed toward the central part of the cloud component at [9, 14.3] km s$^{-1}$ 
(see Figure~\ref{fg5}c). 
It implies the presence of an intensity-depression in the molecular cloud at [9, 14.3] km s$^{-1}$, 
which is referred to as ``cavity" feature. 
The BPN and the EFS are associated with the cloud component at [15.3, 23.3] km s$^{-1}$. 
We find a spatial match between the cloud at [15.3, 23.3] km s$^{-1}$ and the ``cavity" present in the cloud at [9, 14.3] km s$^{-1}$ (see Figure~\ref{fg5}c). 
In addition, some overlapping zones of these two different clouds are also indicated in Figure~\ref{fg5}c (see arrows in Figure~\ref{fg5}c). 

To further explore the broad-bridge feature, we have generated the observed $^{13}$CO profile toward an area marked in Figure~\ref{ufg5x}b 
(see a solid box in Figure~\ref{ufg5x}b). The selected area is seen in the direction of both the clouds. 
The observed $^{13}$CO profile is shown in Figure~\ref{fg5}d, and is computed by averaging the selected area.
The spectrum has almost flattened profile between two velocity peaks around 11.5 and 18.5 km s$^{-1}$, further indicating the presence of the bridge feature. In this context, new molecular line data with better sensitivity and/or resolution will be further helpful.

A detailed discussions on all these results are presented in Section~\ref{sec:disc}.
\subsection{Young stellar populations}
\label{subsec:phot1}
\begin{figure*}
\epsscale{0.8}
\plotone{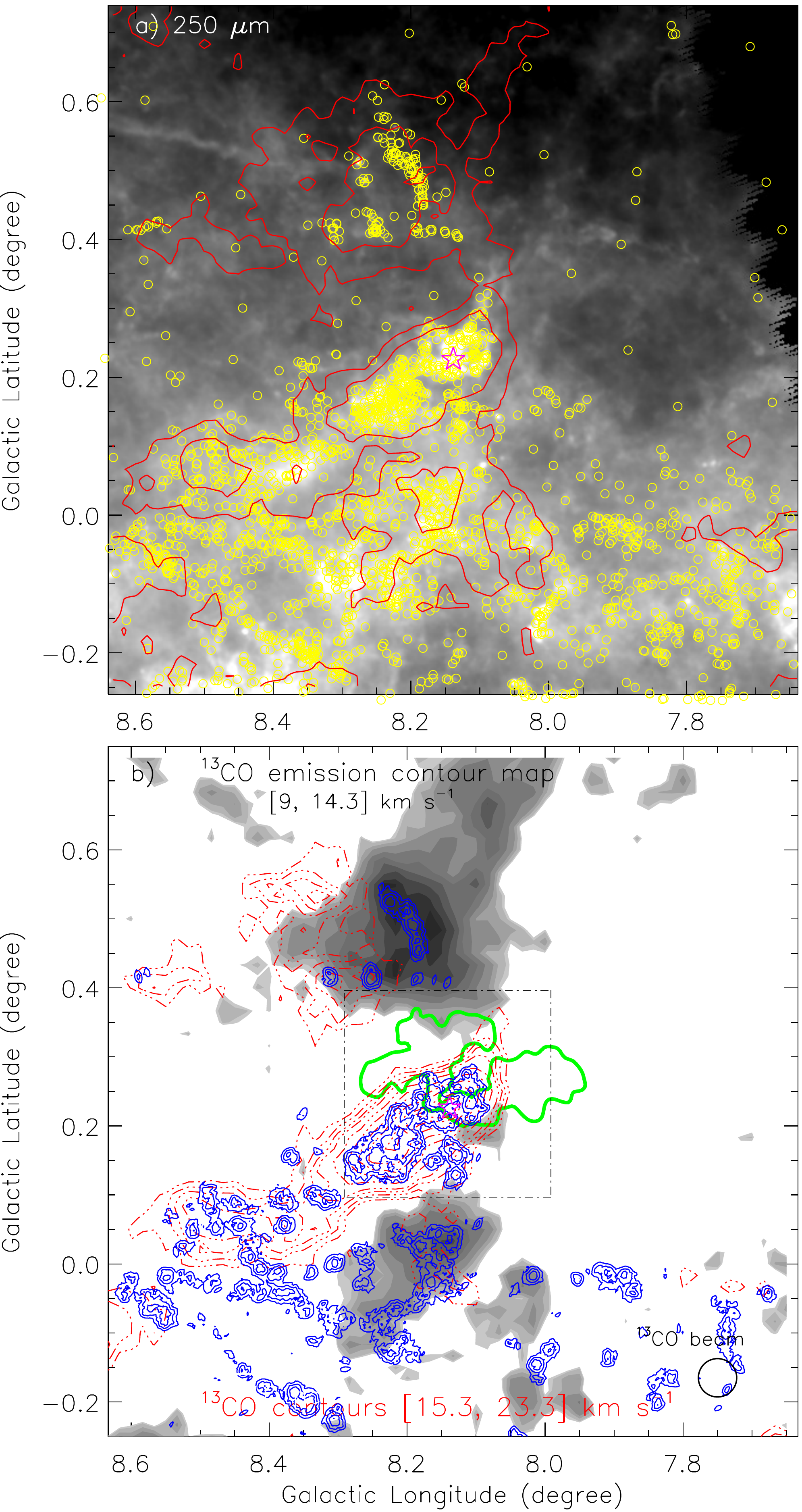}
\caption{a) Overlay of sources with H$-$K $>$ 2.2 mag (see yellow circles) on the {\it Herschel} 250 $\mu$m image. 
The $^{13}$CO emission contours at [8, 25] km s$^{-1}$ are also overlaid on the {\it Herschel} image, and 
are shown with the levels of 10.35 K km s$^{-1}$ (4.5$\sigma$) and 16.66 K km s$^{-1}$ (7.2$\sigma$). 
b) Overlay of surface density contours of sources with H$-$K $>$ 2.2 mag (see yellow circles in Figure~\ref{fg8}a) on the integrated intensity maps of $^{13}$CO (at [9, 14.3] and [15.3, 23.3] km s$^{-1}$). The molecular maps are the same as in Figure~\ref{fg5}c. 
The surface density contours (in blue) are plotted with the levels of 3, 5, and 10 YSOs pc$^{-2}$.  
The {\it Herschel} temperature emission at 23~K is shown with a thick curve (in green; see also Figure~\ref{fg3}a). 
In each panel, a star symbol (in magenta) indicates the position of the 6.7 GHz maser.} 
\label{fg8}
\end{figure*}

In Paper~I and Paper~II, the studies concerning the selection of YSOs were performed for an area highlighted by a solid box in 
Figure~\ref{fg1}b. In this paper, we have identified YSOs for a much wider area around the site G8.14+0.23 (see Figure~\ref{fg1}a). 
To examine the sources with infrared-excess emission, we have used the NIR 
color-magnitude plot (H$-$K/K), as adopted in Paper~II. 
The reliable NIR HK photometric magnitudes of point-like sources in our 
selected target region were obtained from the UKIRT Infrared Deep Sky Survey (UKIDSS) 6$^{th}$ archival 
data release (UKIDSSDR6plus) of the Galactic Plane Survey (GPS) and 
2MASS catalog \citep[see Paper~II for more details and also][]{dewangan15b}. In Paper~II, a color 
condition \citep[i.e., H$-$K $>$ 2.2 mag (or A$_{V}$ = 35.5 mag);][]{indebetouw05} was selected 
to identify a population of sources with infrared-excess emission. 
To obtain this condition, the color-magnitude analysis of sources located in a nearby control field was carried out 
(see Paper~II). 
In our selected target area, we 
find 2150 deeply embedded sources with H$-$K $>$ 2.2 mag, which are overlaid on the {\it Herschel} 
250 $\mu$m and molecular maps (see Figure~\ref{fg8}a). The $^{13}$CO emission contours 
at [8, 25] km s$^{-1}$ are also superimposed on the {\it Herschel} image. 
In Figure~\ref{fg8}b, we have also shown surface density contours of these embedded sources, which are overlaid on the 
molecular maps at [9, 14.3] and [15.3, 23.3] km s$^{-1}$. 
The nearest-neighbour (NN) technique is adopted to derive the surface density contours of selected YSOs \citep[see][for more details]{casertano85,gutermuth09,bressert10,dewangan15a,dewangan15b,dewangan17,dewangan17a}. Following the work of \citet{dewangan15a}, we have generated the surface density map of all the selected embedded sources using a 10$\arcsec$ grid and 6 NN at a distance of 3.0 kpc. 
To infer embedded clusters in star-forming regions, \citet{lada03} considered the minimum surface density of 3 YSO pc$^{-2}$. 
Hence, in Figure~\ref{fg8}b, the surface density contours of YSOs are displayed with the levels of 3, 5, and 
10 YSOs pc$^{-2}$, enabling us to depict the groups of YSOs \citep[e.g.,][]{lada03,bressert10}. 
The groups of YSOs are seen toward both the cloud components at [9, 14.3] and [15.3, 23.3] km s$^{-1}$, revealing a spatial correlation between embedded clusters and molecular gas. This analysis helps us to depict the star-forming zones in the extended molecular clouds, suggesting 
the ongoing star formation activities in both the clouds. 

\begin{figure*}
\epsscale{0.82}
\plotone{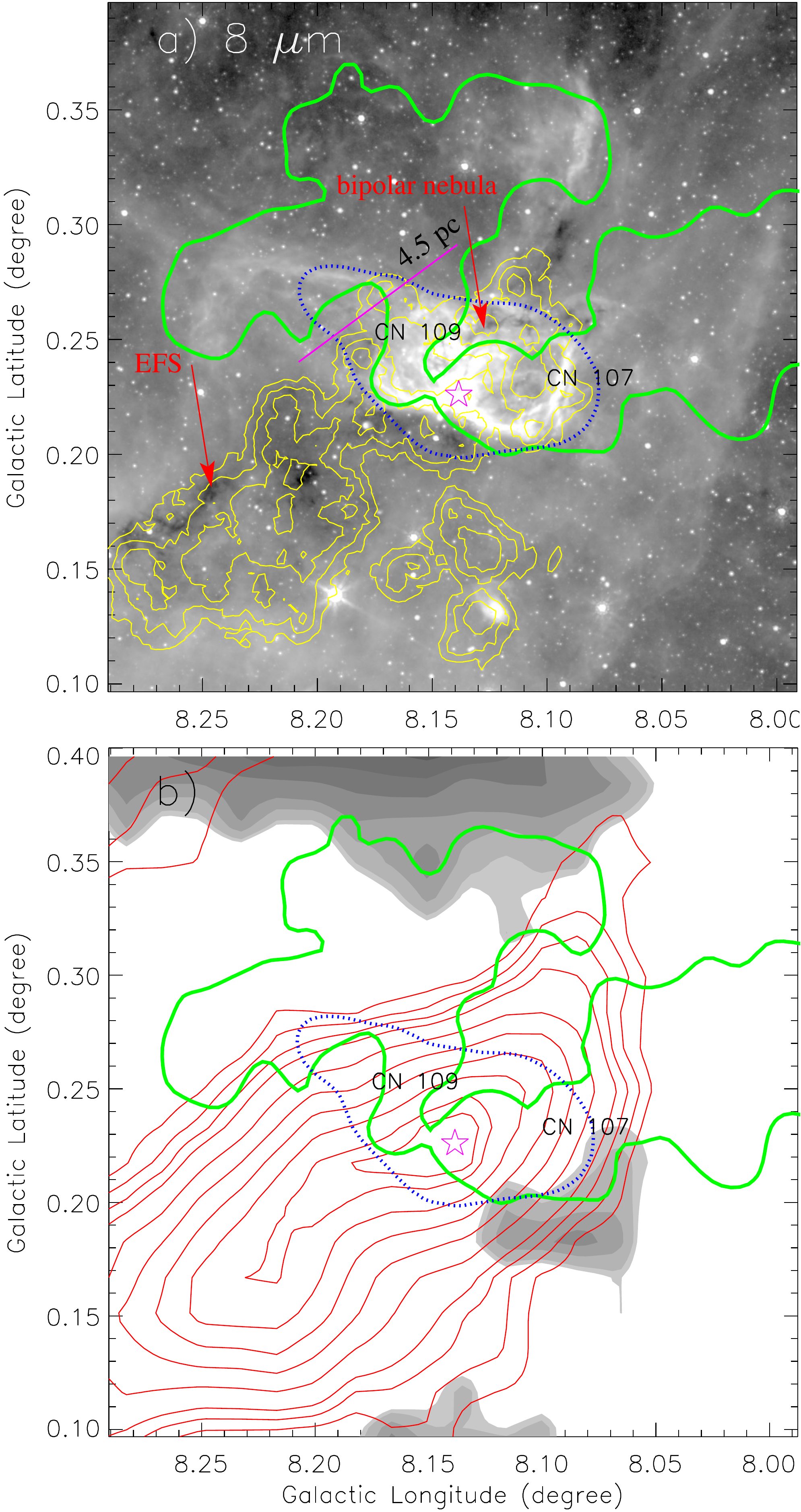}
\caption{A zoomed-in view of the G8.14+0.23 H\,{\sc ii} region (see a broken box in Figure~\ref{fg8}b).
a) Overlay of surface density contours of sources (in yellow) on the {\it Spitzer} 8 $\mu$m image (see also Figure~\ref{fg8}b). The surface density contours are the same as in Figure~\ref{fg8}b. 
b) Spatial distribution of two clouds at [9, 14.3] and [15.3, 23.3] km s$^{-1}$ (see a broken box in Figure~\ref{fg8}b). 
The molecular emission contours are the same as in Figure~\ref{fg8}b. 
In each panel, a star symbol (in magenta) indicates the position of the 6.7 GHz maser.
In both the panels, the NVSS 1.4 GHz radio continuum emission contour (in blue) and 
the {\it Herschel} temperature contour at 23~K (in green) are also drawn.} 
\label{fg9}
\end{figure*}

Figure~\ref{fg9}a shows a zoomed-in view of an area containing the EFS and the BPN. In Figure~\ref{fg9}a, the surface density contours are superimposed on the 
{\it Spitzer} 8.0 $\mu$m image (see also Figure~\ref{fg8}b).
With the help of the surface density analysis, the clusters of YSOs are spatially seen toward the EFS (see also Paper~II). 
The signatures of star formation are also observed toward the edges and the waist of the BPN (see also Paper~I and Paper~II). 
In Figure~\ref{fg9}b, we present the distribution of the clouds at [9, 14.3] and [15.3, 23.3] km s$^{-1}$ in the direction of the same area as shown in Figure~\ref{fg9}a. 
The background contour map represents the cloud at [9, 14.3] km s$^{-1}$, 
while the red contours refer to the cloud at [15.3, 23.3] km s$^{-1}$.  
The NVSS radio continuum contour and the temperature contour at 23 K are also shown in both Figures~\ref{fg9}a and~\ref{fg9}b.
It seems that the star formation activities and the feedback of the massive star may have heated the surroundings, which could be responsible for the extended temperature feature (see also Section~\ref{subsec:temp}). Altogether from Figures~\ref{fg9}a and~\ref{fg9}b, the clusters of YSOs are observed toward the spatial matching zone 
of the elongated redshifted cloud at [15.3, 23.3] km s$^{-1}$ and the ``cavity" feature 
depicted in the cloud at [9, 14.3] km s$^{-1}$, where the exciting star of the G8.14+0.23 H\,{\sc ii} region and the 6.7 GHz maser are investigated. 
\section{Discussion}
\label{sec:disc}
In Paper~I, the bipolar morphology, as the most prominent structure in the 
infrared images, was investigated in the G8.14+0.23 H\,{\sc ii} region, 
which was also found to be part of the EFS (see Paper~II). 
In Paper~I and Paper~II, the star formation in the 
periphery of the BPN was explained by the ionizing feedback of the massive star. 
In Section~\ref{subsec:dstmas}, we have reported the indicative value of the dynamical age of the G8.14+0.23 H\,{\sc ii} region 
to be $\sim$0.3(1.1) Myr for n$_{0}$ of 10$^{3}$(10$^{4}$) cm$^{-3}$. 
Earlier, \citet{evans09} reported a mean age of YSOs to be $\sim$0.44--2 Myr. 
Hence, the formation of the clusters of YSOs seen toward the edges of the BPN could be influenced by the positive feedback of the O-type star (see also Paper~I and Paper~II). 
However, in Paper~II, it was also suggested that the observed clusters of YSOs in the direction of 
the EFS were not triggered by the impact of the O-type star, and might have spontaneously formed. 
This argument was suggested based on the estimation of the pressure of the H\,{\sc ii} region, which was found to be very low (see Paper~II for more details). 
It implies that the birth of the O-type star is not well understood, which is the main goal of this observational work. In this context, we have performed a careful observational study of the molecular gas distribution for $\sim$1$\degr$ $\times$ 1$\degr$ area around the site G8.14+0.23, which has not been conducted before for the selected target (see Section~\ref{sec:coem}). 

\citet{habe92} and \citet{torii15} performed numerical simulations of head-on collisions of two dissimilar clouds to study the CCC process. 
In their simulations, gravitationally unstable cores/clumps at the interface of the clouds were produced by the effect of their compression. 
A cavity was also created in the large cloud, and massive stars were formed at the interface of clouds. After the birth of massive stars, their strong UV radiation produced an H\,{\sc ii} region. 
Subsequently, the cavity in the large cloud was filled with the ionized emission powered by massive stars \citep{torii15}. \citet{inoue13} carried out magnetohydrodynamical numerical simulations of the CCC process, and studied the physical conditions of the interface layer. They found that the collision process significantly enhances the sound speed and the mass accretion rate. 
It was also reported that in the direction of the interface layer, the O-type star can be formed within $\sim$0.1 Myr at a mass accretion rate of 10$^{-4}$ $M_\odot$ yr$^{-1}$ \citep[see also][]{fukui18a}.  
In a very recent simulations, \citet{whitworth18} found the formation of a bipolar H\,{\sc ii} region via the collisions of two clouds. 
In the simulations, a star cluster containing ionizing stars can be formed near the center of the shock-compressed layer. 
Then, the H\,{\sc ii} region powered by the ionizing stars can expand rapidly in the directions perpendicular 
to the layer, leading to the formation of a bipolar H\,{\sc ii} region. 
Hence, it is expected that massive stars and young stellar clusters can be produced at the waist of the bipolar H\,{\sc ii} region or the shock-compressed layer 
\citep[see also][]{inoue13}, where one can obtain high column density materials ($\geq$ 10$^{22}$ cm$^{-2}$).  

Several observational works on the CCC process are also available in the literature \citep[e.g.,][]{loren76,miyawaki86,miyawaki09,hasegawa94,sato00,okumura01,furukawa09,ohama10,ohama17u,ohama17b,ohama17,shimoikura11,torii11,torii15,torii17,torii17b,torii17x,torii17c,nakamura12,nakamura14,shimoikura13,dobashi14,fukui14,fukui15,fukui16,
fukui17,fukui18a,fukui18,fukui18c,fukui18d,haworth15a,
haworth15b,tsuboi15,baug16,dewangan17,dewangan17b,dewangan17c,
dewangan18a,dewangan18b,fujita17,fujita18,gong17,liu17,nishimura17,nishimura18,saigo17,sano17,sano18,tsutsumi17,enokiya18,hayashi18,kohno18a,kohno18b,li18,tachihara18,takahira18,tsuge18}, where one can find observational signposts of the CCC process. As highlighted previously, one expects a spatial and velocity connection of two clouds in the CCC process \citep[e.g.][]{torii17}. In the position-velocity map, the link of blueshifted and redshifted clouds via a lower intensity intermediate velocity emission (i.e. bridge feature) illustrates the velocity connection \citep[e.g.,][]{haworth15a,haworth15b}. 
As mentioned earlier, the bridge feature might trace a compressed layer of gas due to the collision of the clouds \citep[e.g.,][]{torii17}. One can also find the presence of a complementary spatial distribution of clouds in the CCC site, 
which is related to the spatial fit of ``Keyhole/intensity-depression" and ``Key/intensity-enhancement" features \citep[e.g.,][]{torii17,fukui18a}. 
In the CCC process, after the collision of two clouds, 
massive stars and clusters of YSOs can be seen in the shocked-compressed layer of the clouds, and 
a cavity filled with the ionized emission excited by massive stars can be found in one of the clouds.  

\begin{figure*}
\epsscale{1.15}
\plotone{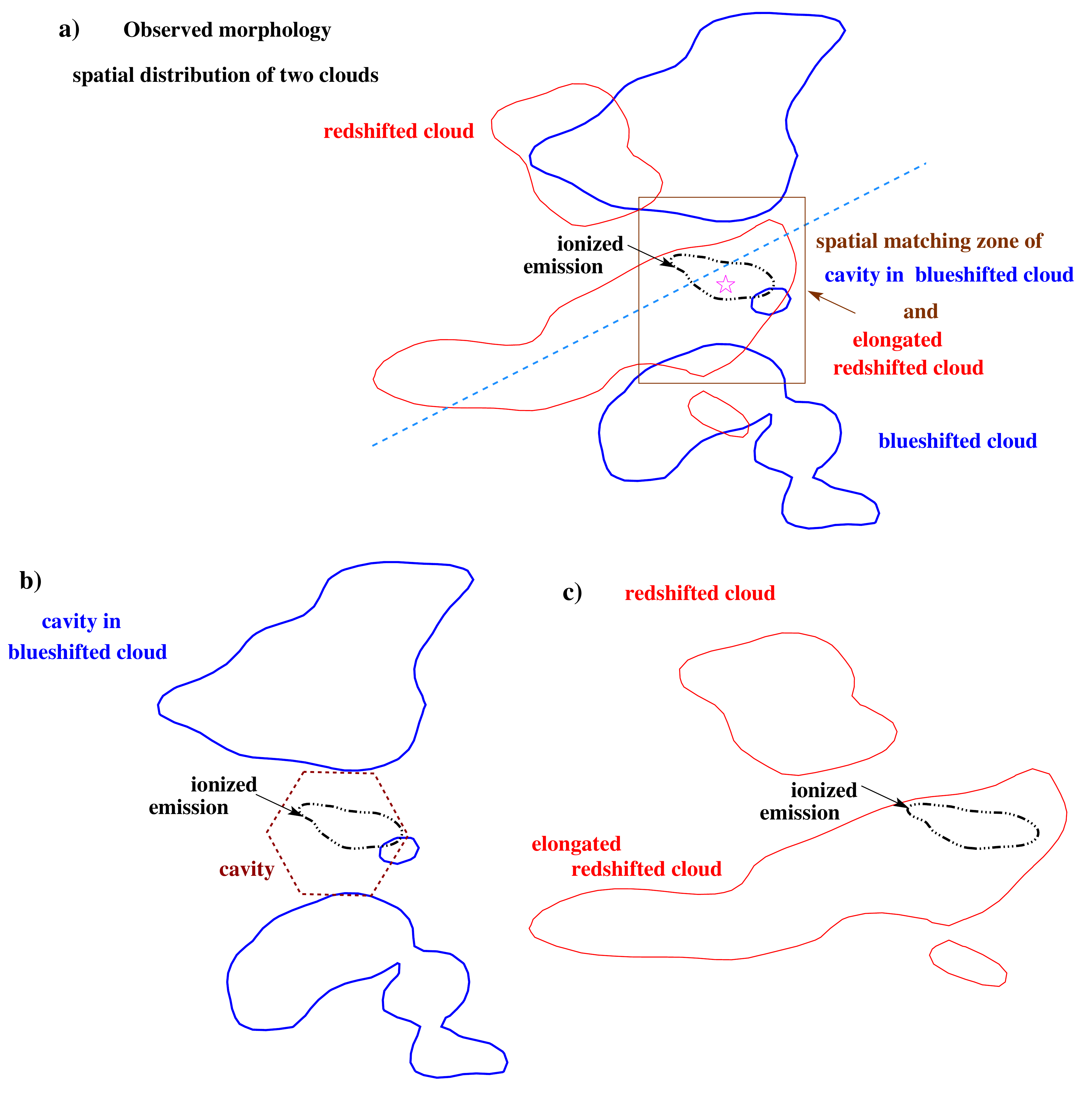}
\caption{a) The panel displays the spatial distribution of redshifted and blueshifted clouds. 
A dotted-dashed contour (in black) represents the location of the H\,{\sc ii} region. 
A star symbol (in magenta) indicates the position of the 6.7 GHz maser. 
A dashed line represents the Galactic eastern-southern to northern-western direction.
b) The panel shows the existence of a cavity (or intensity-depression) in the blueshifted cloud. 
c) The panel highlights the H\,{\sc ii} region in the redshifted cloud.}
\label{xff9}
\end{figure*}

Keeping in mind these observational features of the CCC, 
at least two molecular cloud components (around 11.5 and 18.5 km s$^{-1}$) 
are investigated in the direction of the site G8.14+0.23 (see Figures~\ref{ufg5x}b,~\ref{fg4},~\ref{fg5}a, and~\ref{fg5}b). 
Various observed components related to the site G8.14+0.23 are summarized in Figure~\ref{xff9}a. 
This figure illustrates a complementary spatial distribution of clouds 
(i.e., spatial fit between the cavity in the blueshifted cloud and the elongated redshifted cloud; see a solid box in Figure~\ref{xff9}a). 
The cavity or intensity-depression in the blueshifted cloud is highlighted by a broken hexagonal (see Figure~\ref{xff9}b), 
while the redshifted cloud is shown in Figure~\ref{xff9}c. 
In the position-velocity space, the signature of the bridge feature between the two clouds 
is also found (see Section~\ref{sec:coem} and also Figure~\ref{fg5}). 
An almost V-shaped velocity structure is also seen in the position-velocity map (see Figure~\ref{fg5}a), which can be originated in the position-velocity space due to the combination of the intensity-depression/cavity and the bridge feature \citep[e.g.,][]{anathpindika10,fukui18a,fukui18c}. 
High-resolution and better sensitivity molecular line data will be needed to gain more information regarding the V-shaped velocity structure. 
Altogether these observational facts are consistent with the signatures of the CCC scenario.
Therefore, it appears that the observed molecular morphology hints the post-collision condition in the site G8.14+0.23 (see Figure~\ref{xff9}a). The ``elongated redshifted cloud" is analogous to the shock-compressed layer as 
seen in the simulations of the CCC process. The shock-compressed layer is expected at the location of 
impact in the post-collision condition, whereas the ``cavity" in the blueshifted 
cloud denotes the collision site \citep[e.g.,][]{torii17}. Hence, before the collision process, no cavity was predicted in the blueshifted cloud. 
In the direction of the BPN, the column densities are observed in the range of $\sim$1.8--3.5 $\times$ 10$^{22}$ cm$^{-2}$ ($A_V$ $\sim$19--37 mag). 
Furthermore, the present data do not allow to obtain the exact information of the viewing angle of the collision 
to the line of sight. 
With the knowledge of the distribution of the cavity and the elongated redshifted cloud, 
the collision of two clouds might had taken place in the Galactic northern-western direction (see a broken line in Figure~\ref{xff9}a), where the BPN is seen. 
Hence, the broken line may indicate the direction of the relative motion of the two clouds (see Figure~\ref{xff9}a). 

Note that the signposts of massive stars (i.e., H\,{\sc ii} region and 6.7 GHz methanol maser), the embedded clusters, the BPN, and the extended temperature feature are 
observed toward the spatial matching zone of the cavity and the redshifted cloud/shock-compressed layer (see Figure~\ref{fg9} and also a solid box in Figure~\ref{xff9}a). 
In order to infer the impact of the CCC, the information of the collision time-scale is essential. 
Following the equation given in \citet{mckee07}, one can estimate the 
collision time-scale using the relative velocity of the two clouds, collision length-scale, 
and ratio of number densities of post-collision and pre-collision regions \citep[see also equation~2 in][]{henshaw13}. 
The time-scale of the accumulation of material at the collision points or the collision 
time-scale \citep{mckee07,henshaw13} can be computed using the expression below:
\begin{equation}
t_{\rm accum} = 2.0\,\bigg(\frac{\rm l_{ccc}}{0.5\,{\rm pc}} \bigg) \bigg(\frac{v_{\rm
rel}}{5{\rm \,km\,s^{-1}}}\bigg)^{-1}\bigg(\frac{n_{\rm pstccc}/n_{\rm
preccc}}{10}\bigg)\,{\rm Myr} 
\end{equation}
where, l$_{ccc}$ is the collision length-scale, ${v_{\rm
rel}}$ is the relative velocity of clouds, n$_{preccc}$ is the mean 
density of pre-collision region, and n$_{pstccc}$ is the mean density of post-collision region.  
In the calculation, we have assumed a viewing angle (= 45$\degr$) of the collision to the line of sight. 
In Section~\ref{subsec:dstmas}, we have used the size of the H\,{\sc ii} region (i.e. 2 $\times$ R$_{HII}$) to be $\sim$4.5 pc, where the ionized emission is spatially distributed (see Figure~\ref{fg9}a). 
Hence, we have used the size of the H\,{\sc ii} region to estimate the collision length-scale (l$_{ccc}$), 
which is adopted to be $\sim$6.4 pc (= 4.5 pc/sin(45$\degr$)). 
In the direction of the site G8.14+0.23, the velocity separation of the two clouds is $\sim$7 km s$^{-1}$. 
It gives the observed relative velocity (${v_{\rm
rel}}$) to be $\sim$10 km s$^{-1}$ (= 7 km s$^{-1}$/cos(45$\degr$)). 
Note that in this paper, the existing observational data sets do not allow us to determine the exact ratio of the mean densities of the pre- and post-collision regions. 
In the collision process, one can expect the increase of density in the post-collision region against the pre-collision site.
Hence, the density ratio is expected to be larger than unity (i.e. n$_{pstccc}$/n$_{preccc}$ $>$ 1). 
If we use a range of the ratios of densities (i.e., 1--10) in equation~2 then the collision timescale range is 
estimated to be $\sim$1.28--12.8 Myr.  

In Section~\ref{subsec:dstmas}, the dynamical age of the G8.14+0.23 H\,{\sc ii} region is computed to be $\sim$0.3(1.1) Myr 
for the ambient density of 10$^{3}$(10$^{4}$) cm$^{-3}$. In this analysis, the spherical morphology of the H\,{\sc ii} region was assumed, which 
evolved into the homogeneous medium \citep[see][for more details]{dyson80}. 
Both these assumptions are unlikely for the G8.14+0.23 H\,{\sc ii} region. 
Therefore, the dynamical age of the H\,{\sc ii} region as well as the collision timescale can be considered as indicative values. Despite the uncertain nature of these calculations, the ratio of the dynamical age of the H\,{\sc ii} region and the timescale for mass build up 
during the collision process appears to be less than unity (i.e., t$_{dyn}$/t$_{\rm accum}$ $<$ 1), which indicates that the proposed hypothesis is plausible.  
Additionally, in the site, the early phase of massive star formation ($<$ 0.1 Myr) 
has also been reported through the detection of the 6.7 GHz methanol maser (see Paper~II). In the numerical simulations, the O-type star is formed within a few times of 10$^{5}$ yr \citep{inoue13}.   
It appears that the CCC process began before the formation of massive stars in our selected site.   
Hence, the birth of the O-type star located in the waist of the BPN 
seems to be triggered by the CCC process \citep[see also the simulation of][]{whitworth18}. 
Note that star formation activities toward the edges of the BPN could also be triggered by the O-type star, which we 
cannot entirely rule out.   

Our case study of the bipolar H\,{\sc ii} region G8.14+0.23 encourages us to examine the CCC process in other known sample of 
Galactic bipolar H\,{\sc ii} regions \citep[e.g.,][]{churchwell06,churchwell07}, which will further allow us to observationally 
assess the existing theoretical simulations concerning the triggered O-type star formation on the waist of the bipolar H\,{\sc ii} regions.   
\section{Summary and Conclusions}
\label{sec:conc}
To understand the formation of massive stars and young stellar clusters, a large-scale area of the 
site G8.14+0.23 (size $\sim$1$\degr$ $\times$ 1$\degr$) is investigated using a multi-wavelength approach.
The major outcomes of this paper are as follows:\\
$\bullet$ The G8.14+0.23 H\,{\sc ii} region, having a dynamical age of $\sim$0.3(1.1) Myr for n$_{0}$ of 10$^{3}$(10$^{4}$) cm$^{-3}$, is powered by at least an O-type star, which is observed toward the waist of the BPN.\\
$\bullet$ The NANTEN2 $^{13}$CO line data trace molecular gas associated with the site G8.14+0.23 in a velocity range of [8, 25] 
km s$^{-1}$. Furthermore, two extended molecular cloud components at [9, 14.3] and [15.3, 23.3] km s$^{-1}$ are unveiled toward 
the selected site. The data have also shown their physical connection in both space and velocity.\\
$\bullet$ A spatial complementary distribution concerning the collisions of two clouds is found in the selected site. 
A ``cavity/intensity-depression" feature is evident in the blueshifted cloud (i.e., [9, 14.3] km s$^{-1}$), 
and is spatially matched by the elongated redshifted cloud (i.e., [15.3, 23.3] km s$^{-1}$). 
A broad-bridge feature is also identified in the position-velocity space.\\ 
$\bullet$ With the help of deep NIR photometric data of point-like sources, a total of 2150 embedded sources 
with H$-$K $>$ 2.2 mag are identified in the selected target field. Groups of the embedded infrared-excess sources 
are observed in both the clouds (including the EFS and the BPN). \\
$\bullet$ The {\it Herschel} temperature map shows an extended temperature structure (T$_{d}$ = 23--36 K) in the direction of 
the G8.14+0.23 H\,{\sc ii} region.\\
$\bullet$ Signposts of massive star formation (i.e., 6.7 GHz methanol maser and H\,{\sc ii} region), the BPN, the embedded clusters, and the extended temperature structure are found toward the spatial matching zone of the cavity and the redshifted cloud. 
In the direction of this matching zone, the column densities are found to 
be $\sim$1.8--3.5 $\times$ 10$^{22}$ cm$^{-2}$ (or A$_{V}$ $\sim$19--37 mag).\\ 
$\bullet$ The outcomes of the present work suggest the interaction of molecular clouds in the site G8.14+0.23.\\ 

Considering our new observational findings in this paper, we suggest that the CCC scenario can explain the star formation history in the site G8.14+0.23. Our observed results are in 
qualitative agreement with the proposed numerical simulations of the CCC. 
In this connection, new molecular line data with better sensitivity and high-resolution will be further useful.
\acknowledgments 
We thank the anonymous reviewer for a critical reading of the manuscript and several useful comments and 
suggestions, which greatly improved the scientific contents of the paper. 
The research work at Physical Research Laboratory is funded by the Department of Space, Government of India. 
This work is based on data obtained as part of the UKIRT Infrared Deep Sky Survey. This publication 
made use of data products from the Two Micron All Sky Survey (a joint project of the University of Massachusetts and 
the Infrared Processing and Analysis Center / California Institute of Technology, funded by NASA and NSF), archival 
data obtained with the {\it Spitzer} Space Telescope (operated by the Jet Propulsion Laboratory, California Institute 
of Technology under a contract with NASA). 
NANTEN2 is an international collaboration of 11 universities: Nagoya University, Osaka Prefecture University, 
University of Bonn, University of Cologne, Seoul National University, University of Chile, University of 
New South Wales, Macquarie University, University of Sydney, University of Adelaide, and University of ETH Zurich.
\end{document}